\pdfoutput=1

\documentclass{tcibook}
\usepackage{fancyhea}
\usepackage{work}
\usepackage{bm}       
\usepackage{graphicx}

\newcommand{\nc}{\newcommand}  

\def\beq{\begin{equation}}
\def\eeq#1{\label{#1}\end{equation}}
\def\eeqn{\end{equation}}

\newenvironment{Eqnarray}%
   {\arraycolsep 0.14em\begin{eqnarray}}{\end{eqnarray}}
\def\beqa{\begin{Eqnarray}}
\def\eeqa#1{\label{#1}\end{Eqnarray}}
\def\eeqan{\end{Eqnarray}}

\nc{\ra}{\rightarrow}  
\nc{\slsh}{\slash\hspace*{-0.22cm}}
\def\Re{{\cal R \mskip-4mu \lower.1ex \hbox{\it e}\,}}
\def\Im{{\cal I \mskip-5mu \lower.1ex \hbox{\it m}\,}}

\nc{\vev}[1]{ \left\langle {#1} \right\rangle }
\nc{\bra}[1]{ \langle {#1} | }
\nc{\ket}[1]{ | {#1} \rangle }
\nc{\fb}{\,{\rm fb}^{-1}}
\nc{\ev}{{\rm eV}}
\nc{\kev}{{\rm keV}}
\nc{\Mev}{{\rm MeV}}
\nc{\gev}{{\rm GeV}}
\nc{\tev}{{\rm TeV}}
\nc{\mev}{{\rm MeV}}

\def\del{\partial}
\def\Dslash{\not{\hbox{\kern-4pt $D$}}}
\def\dslash{\not{\hbox{\kern-2pt $\del$}}}
\def\pslash{\not{\hbox{\kern-2pt $p$}}}
\def\ETmiss{ \not{\hbox{\kern-4pt $E$}}_T }

\def\msb{{\bar{\ssstyle M \kern -1pt S}}}

\begin{document}

\def\bibname{References}
\bibliographystyle{plain}

\raggedbottom

\pagenumbering{roman}

\parindent=0pt
\parskip=8pt
\setlength{\evensidemargin}{0pt}
\setlength{\oddsidemargin}{0pt}
\setlength{\marginparsep}{0.0in}
\setlength{\marginparwidth}{0.0in}
\marginparpush=0pt


\pagenumbering{arabic}

\renewcommand{\chapname}{chap:intro_}
\renewcommand{\chapterdir}{.}
\renewcommand{\arraystretch}{1.25}
\addtolength{\arraycolsep}{-3pt}


 
\chapter{Snowmass 2013 Computing Frontier Storage and Data Management}
\label{chap:mag}

\begin{center}\begin{boldmath}

\begin{center}
{\large Conveners: Michelle Butler$^1$,  Richard Mount$^2$\\
  Observer: Mike Hildreth$^3$} \\
\bigskip
$^1${\it National Center for Supercomputing Applications, University of Illinois at Urbana Champaign}\\
$^2${\it Particle Physics and Astrophysics, SLAC National Accelerator Laboratory}\\
$^3${\it Department of Physics, University of Notre Dame }\\
\end{center}


\end{boldmath}\end{center}


\section{Input from the Science Frontiers}
\label{sec:cpfi5-input}
\subsection{Energy Frontier}
Experiments at energy frontier hadron colliders already generate over a petabyte per 
second of data at the detector device level.  Triggering and real-time event filtering 
is used to reduce this by six orders of magnitude for a final rate to persistent storage 
of around one gigabyte per second in the case of LHC experiments at the start of Run 2.  
The cost of storing and analyzing data is a major fraction of the operational cost of LHC experiments, and thus the rate to persistent storage is determined by physics requirements and, perhaps even more, by what is affordable.

Science at energy frontier lepton colliders is unlikely to be constrained by data 
management and storage issues.

The current practice of ATLAS and CMS is to treat the all the data written to persistent 
storage equally through the production phases of reconstruction, re-reconstruction and 
worldwide distribution of a complete set of data ready for physics analysis.  Flagging 
a large fraction of the persistent data for storage on tape only, with no further 
reconstruction or distribution unless a physics case arose, would cut costs or allow 
the rate to persistent storage to be raised.

The vast majority of LHC data storage and data access relates to derived or simulated 
data stored and accessed using ROOT\cite{Antcheva:2009zz} persistency\cite{Canal:2011zz}.  Efficient and agile data access is 
already a major issue that is expected to rise in importance by the time of LHC Run 3.  
Beyond HEP, efficient and agile data access will underpin future successes in data-intensive 
scientific and commercial applications.  It is hard, but not impossible, to believe that HEP 
will continue to be best served by a persistency solution that appears to be confined to HEP.

\begin{figure}[h]
\centering
\includegraphics[width=0.7\textwidth]{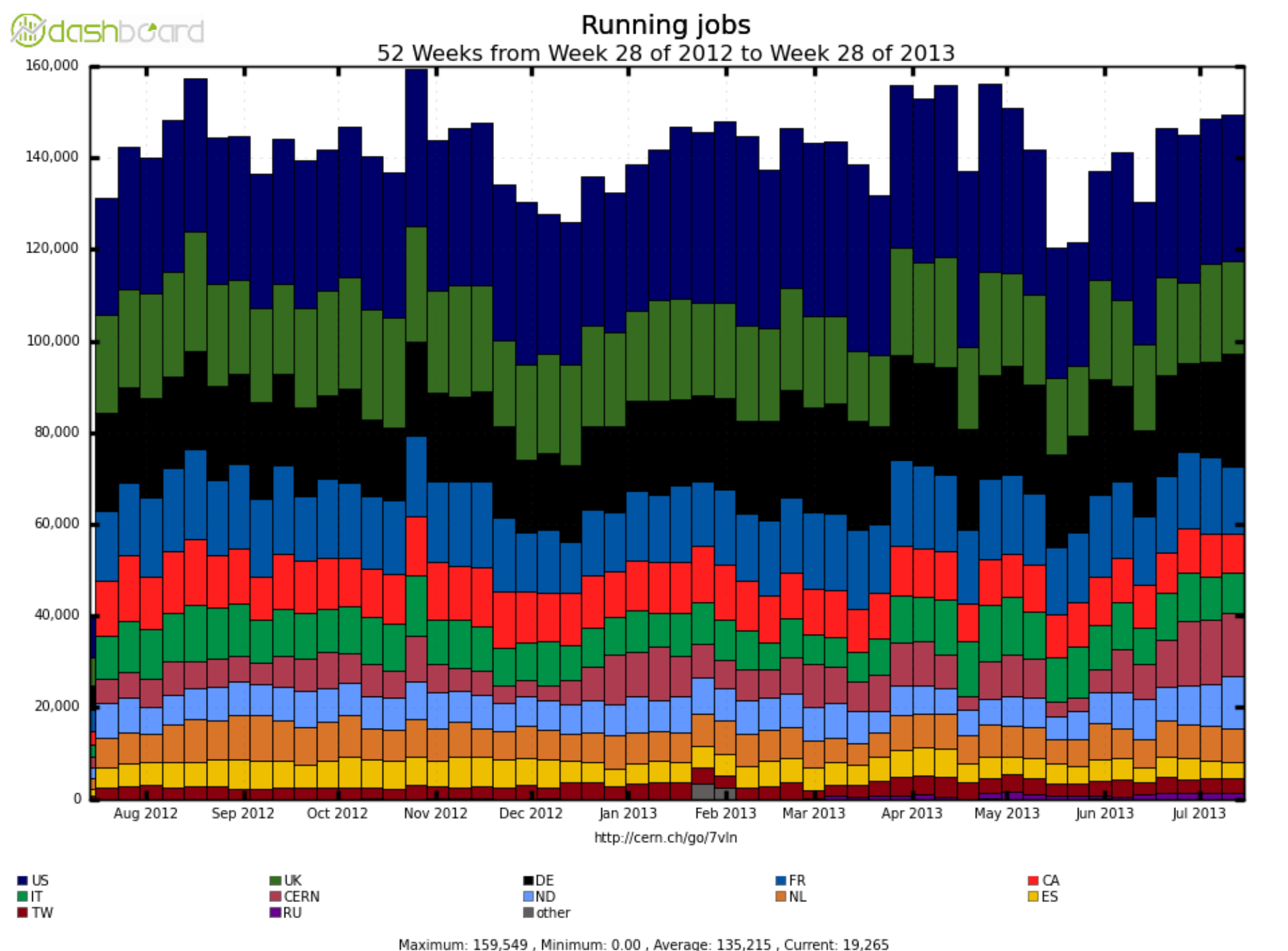}
\caption{ATLAS jobs executing on the WLCG}
\label{fig:atlas-jobs}
\end{figure}

\begin{figure}[h!]
\centering
\includegraphics[width=0.7\textwidth]{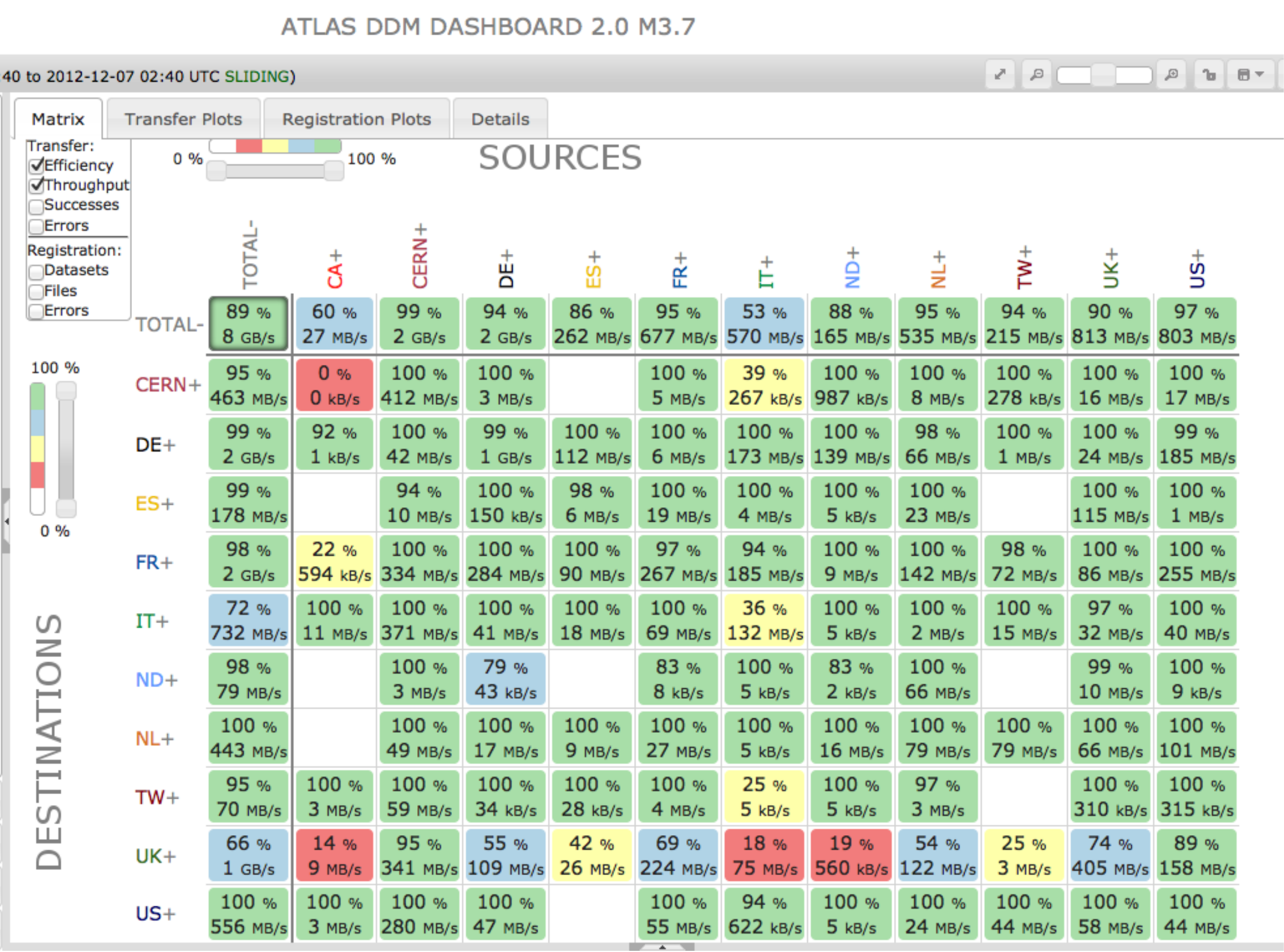}
\caption{ATLAS distributed data management: Top level source-destination matrix showing success rates and data transfer rates}
\label{fig:atlas-ddm}
\end{figure}

\begin{figure}[h!]
\centering
\includegraphics[width=0.7\textwidth]{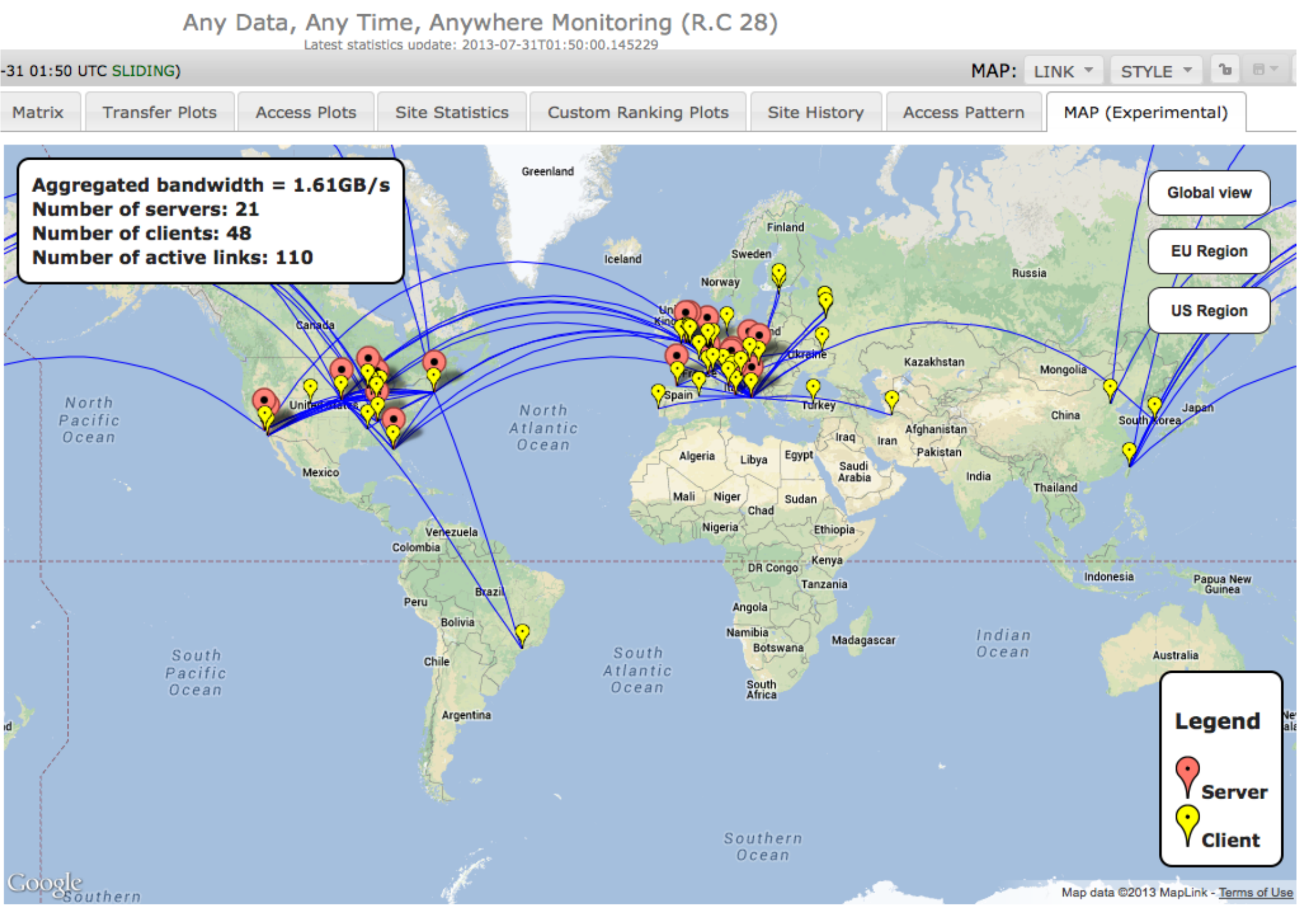}
\caption{Map showing the CMS ``Any data Anywhere Anytime'' distributed direct access system based on xrootd\cite{Bauerdick:2012st} software}
\label{fig:cms-aaa}
\end{figure}

Distributed workflow and data management is proving to be a major success of the LHC experiments, as 
illustrated in Figures \ref{fig:atlas-jobs},\ref{fig:atlas-ddm},\ref{fig:cms-aaa}, and is an area where the US brings much of the intellectual 
leadership.  But this success is costly. For example the US-ATLAS M\&O-funded effort that 
contributes to developing and operating the ATLAS distributed workflow and data management 
system amounts to around \$4.5M per year.  This is in addition to the effort required to operate 
the US Tier 1 and Tier 2 facilities.  It is difficult to imagine that the LHC-focused effort 
could continue at this level for two decades, and even more difficult to imagine that other HEP 
or wider science activities could take advantage of the LHC experiments current achievements 
given their operational cost and complexity. 

With respect to data/physics preservation and open availability, the LHC experiments are actively 
developing appropriate policies. Nascent efforts aimed at data preservation for internal 
re-use and some moves towards public access have begun at the LHC.  The Tevatron experiments 
have larger and more focused activities at this time; BaBar and the HERA experiments have
devloped substantial data preservation infrastructures.

\subsection{Intensity Frontier}
Lepton colliders in intensity frontier ``factory mode'' also run up against the cost of storage, 
but the physics of lepton collisions is relatively clean and recording all events relevant to the 
targeted physics has proved possible in the past and is a realistic expectation for the future.  
The Belle II TDR estimates a data rate to persistent storage of 0.4 to 1.8 gigabytes/s, which is 
comparable to LHC Run-2 rates but without the need to discard data with significant physics content.

Most of the many other intensity frontier experiments do not individually challenge storage 
capabilities, but there is a recognition that data management (and workflow management) is 
often inefficient and burdensome. Most experiments find it hard to escape from the comfort 
and constriction of limiting all their data-intensive work to a single site -- normally Fermilab. 
The statement ``all international efforts would benefit from an ATLAS-like model'' was made and 
should probably be interpreted as a need for ATLAS-like data management functions at a much 
lower cost and complexity than the current ATLAS system.

With respect to data preservation and open availability, the intensity frontier community 
does not have a unified plan.  BaBar has implemented its Long-Term Data Analysis system in 
the form of a stable, secure computing and storage cluster to support active analysis by 
BaBar members for as long as possible. As a whole, the intensity frontier community recognizes 
that the issue of data preservation and open availability exists and is pressing. It was widely 
acknowledged that additional person-power is required to mount an extensive effort in this area.

\subsection{Cosmic Frontier}
The cosmic frontier presents several faces, each presenting its own challenges for data 
management and storage: terrestrial sky survey telescopes, terrestrial radio telescopes, 
HEP-detector-in-space telescopes, and large-scale simulations.

\subsubsection{Sky Surveys}
The Sloan Digital Sky Survey pioneered the use of innovative database technology to make 
its data maximally useful to scientists.  This approach continues with LSST, notably the 
development of a multi-petabyte scalable object catalog database that is capable of rapid 
response to complex queries.  The data management needs of the sky surveys -- handling image 
catalogs and object catalogs -- appear very different from those of experimental HEP, but 
nevertheless, the baseline LSST object catalog employs HEP's xrootd technology in the key 
role of providing a switchyard between MYSQL front ends and thousands of MYSQL backend servers.  
LSST's 3.2 gigapixel camera will produce 15 terabytes per night, building up to over 100 
petabytes of images and 20 petabytes of catalog database during the first ten years.

Although the basic data-access technology to make LSST science achievable has already been 
demonstrated, it is certain that a vigorous LSST science community will want to attempt many 
scientific studies that will be poorly served without major additional developments.  
Not all LSST science will be possible using only the object catalog database. In particular, 
studies such as those for dark energy effects, of particular interest to the HEP community, 
are likely to require reprocessing of the LSST image data on HEP analysis facilities.  
The model for funding and executing these studies is not yet clear.

The Dark Energy Survey (DES) can be considered a precursor to LSST, taking data with a 
0.6 gigapixel camera for five years from 2012 culminating in a petabyte dataset.

CTA has a rather specialized real-time data challenge where some 30 gigabytes/s of data 
must be gathered and processed in real time from about 100 telescopes spread over a square kilometer.

\subsubsection{Terrestrial Radio Telescopes}
Arrays of radio telescopes can present a data-volume challenge comparable with that 
posed by energy frontier hadron collider experiments.  The most extreme example now being 
planned is the European-led Square Kilometre Array (SKA) project that expects to complete 
its Phase I system in 2020. SKA will feed petabytes/s to correlators that will synthesize 
images in real time, producing a reduced persistent dataset on the scale of 300 to 1500 
petabytes per year.  These volumes can only be realized if considerable evolution of 
computing and storage costs happens by the time SKA data flows.  Although SKA currently 
has no US involvement, it presents a concretely planned example of the technologies and 
data-related challenges that will certainly be faced by US scientists involved in projects 
in the same timeframe.

Today's example of the SKA concept is the Murchison Wide-Field array where a raw 
15.8 gigabytes/s is processed to a produce a stored 400 MB/s.

\subsubsection{“HEP Detectors” in Space}
Examples include the Fermi Gamma Space Telescope (FGST) and the Alpha Magnetic 
Spectrometer (AMS-02).  These detectors have front-end data rates far lower than 
LHC experiments and would not be constrained by terrestrial storage and data analysis 
capabilities.  The choke points determining their trigger rates to persistent storage 
are the limited bandwidth of the downlinks that bring data back to Earth.  The 
necessary conservatism applied by NASA and other space agencies to placing new 
technology in space seems guaranteed to keep downlink bandwidths well below rates 
that would make storage and data distribution a challenge in the future.  
Nevertheless, these detectors are built and operated by large collaborations and 
thus require functional distributed data management.

\subsubsection{Simulations}
Simulation provides our only way to perform ``experimental cosmology'' since only one 
universe is observable.  Simulation also plays a vital role in understanding all 
aspects of astrophysics, such as supernovae, for which only very limited observation 
data can be collected for each occurrence. Finally, simulation is needed for the 
design of observational programs and for their detailed technical elements.

Already today, post-processing of simulation data presents a major data-intensive 
computing challenge, requiring data management, large-scale databases and tools for 
data analytics.  Some of today's pain relates to the much more ready availability of 
national resources for computation than those for data management and analysis: ``we 
can easily generate many petabytes from simulations and have [almost] no place to 
store them and analyze them''.

There is some expectation that compute-intensive simulation will be co-located with the 
data-intensive facilities for analysis of the simulation, but powerful, easy to use 
analysis tools will still need to be developed.

\subsubsection{Cosmic Frontier Data Preservation and Access}

The images, and tabular object catalogs of sky surveys and other image-based astronomy 
are readily intelligible by other scientists and even the general public.  

From the 
experimental HEP perspective, data preservation and open availability is relatively 
simple to achieve for image-based astronomy once policies have been decided.  

However, the simple approach to open access will become more challenging for data-intensive image-based activities such as LSST.  Even though the data schema may be intelligible to all, analysis of the data can require massive resources, effectively limiting open access to the images or even to the multi-petabyte object catalog.  Current plans for LSST public access to
data include providing limited computing resources, with data access limited to those individuals or organizations
whose proposals for data analysis have been approved.

The raw persistent data from ``HEP-Detector-like'' devices contrasts markedly with that from the 
image-based telescopes.  Like data from almost any HEP experiment, they are 
intelligible only to a few experts until substantial reconstruction and analysis 
has been performed.  They present the same data preservation and open availability 
challenges as HEP experiments, and may be subject to the higher availability 
expectations typical of image-based astronomy.

\subsection{Lattice Field Theory (LQCD)}
Like other simulation-based sciences, LQCD uses massive simulations on national 
supercomputer facilities, followed by intense analysis of the resultant data.  The LQCD configuration generation step is performed on massively parallel supercomputers 
because these are best adapted to the problem, whereas the subsequent, and by no means less compute-intensive, analysis step is performed on HEP-funded 
throughput-optimized systems because these are the most appropriate for this step.

LQCD has significant, but not problematic data volumes that must be managed and transmitted 
between the two steps, but does not face major data-related challenges.

\subsection{Perturbative QCD}
Data-related challenges are expected to remain minor in relation to those of other branches of HEP.

\subsection{Accelerator Science}
In broad summary, accelerator science is not a driver in data (or networking), but would 
certainly welcome access to the easy-to-use data management and analysis tools that are 
the goal of a wide range of HEP experiments.

Accelerator science has a long held dream of being able to perform predictive simulations 
in close-to-real time so that feedback can be provided to physicists in the control room as 
they strive to optimize accelerator performance.  A likely scenario involves running a massively 
parallel simulation for a relatively short time on a remote Leadership Class Facility, followed 
by the rapid transfer of tens or hundreds of terabytes of simulated data to local facilities 
for rapid analysis.  This scenario is becoming achievable, but will stretch the limits of data 
transmission bandwidths and of rapid data analysis.

\subsection{General Considerations for Large-Scale HEP}
Large and costly experiments or telescopes, and even some simulations, require 
international collaboration.  Whatever could be done by one nation can be done 
on a larger scale with more science reach as a collaborative project.  In such 
international projects, the data storage and analysis can take advantage of funding 
from many nations, access to large shared resources, opportunistic access to a wide 
range of resources, and access to and development of distributed expertise.  
The price to be paid is the complexity of a geographically distributed system.

Thus HEP, and indeed any data-intensive science operating at a comparable 
international scale, needs to use a combination of wide area networks, 
distributed storage, distributed computing, and the software technologies 
to co-manage efficient worldwide work flow and data flow.  It seems clear 
that future HEP does not need many different solutions to this challenge, 
and may have much to gain by identifying commonalities with other 
scientific or even commercial fields.

\section{Technology Outlook}
\label{sec:cpfi5-technology}
\begin{figure}[h]
\centering
\includegraphics[width=0.7\textwidth]{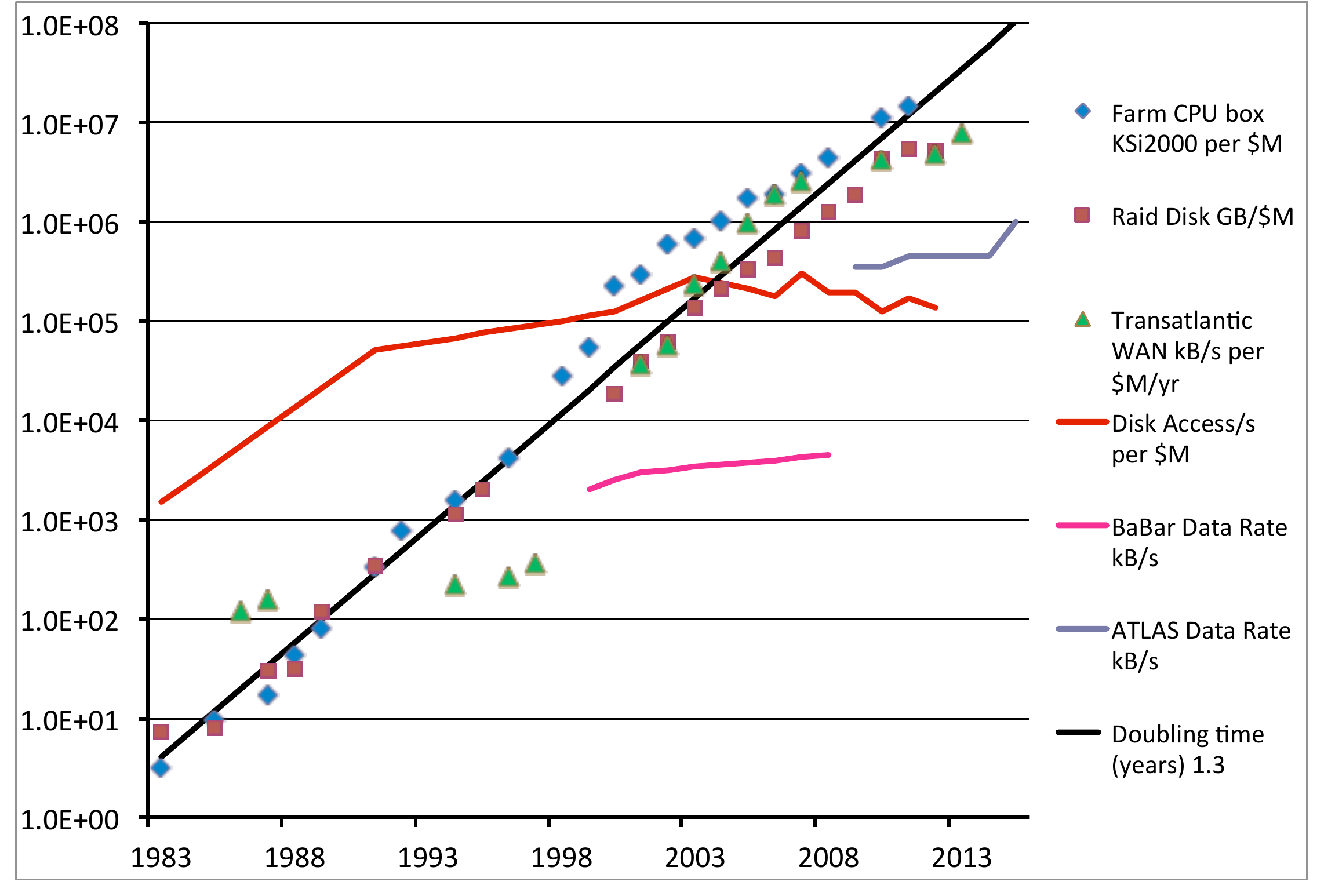}
\caption{Historical trends in technology used for HEP computing.  The data reflect cpu and disk purchases for L3 at CERN, and BaBar and ATLAS at SLAC.  The network data are for the LEP3Net and USLHCNet links managed by Caltech}
\label{fig:technology-evolution}
\end{figure}

A simplistic prediction for the future evolution of technology is that it 
would to continue to 
evolve as it has in the past.  Figure \ref{fig:technology-evolution} shows a highly selective, but relevant, 
view of technology evolution, charting how much you can get for \$1M in disk storage, 
CPU power and long-haul network links that have been bought over three decades 
for some particular experimental HEP activities.  The figure is  selective in that 
it involves real purchases of hardware and services for a limited range of 
experiments -- L3, BaBar, and parts of the LHC program. Some notable features of the 
evolution are:
\begin{itemize}
\item
Over three decades the line ``doubling every 1.3 years'' is a good match to the average 
CPU evolution and is not far from the average disk capacity evolution.
\item
The data do not exclude a marked slow down in evolution from about 2010.
\item
The network evolution shows a major discontinuity, corresponding to de-regulation 
and the end of European PTT monopolies.
\item
Disk accesses per \$M -- almost totally dominated by the unchanging rotational 
speed of disks -- has changed hardly at all in two decades.
\item
Comparison of technology evolution with the BaBar and ATLAS raw data rates to 
persistent storage (just two examples of HEP's ``data frontier'') shows a similar 
distance from the technology evolution and hence, perhaps a similar level of 
technological challenge but separated by more than a decade in time. 
\end{itemize}

\subsection{Storage Futures}
\subsubsection{Tape}
The death of tape storage has been predicted for more than a decade, but today 
its future seems to be more assured than at any time in the last decade.  
Much of this change relates to the problems with disk technology described 
below, but tape storage does have the intrinsic properties of negligible 
power usage, a different set of failure modes and often lower failure rate 
than disks, and persistently lower cost than disks.  The scientific and the 
commercial world has and will continue to have, a need for archival or ``just 
in case we need it'' storage with these properties.

Tape does fail, so HEP has traditionally taken advantage of its enforced distributed 
approach to computing to make a copy or copies elsewhere in the world.  A cheaper 
approach is possible with RAIT (redundant array of individual tapes), analogous 
to RAID, provided that the data are stored away from areas prone to major natural
disasters.

Data volumes in science beyond HEP seem to be doubling every year according to NCSA 
experience.  This makes sense as driven not just by the increasing data hunger of 
individual sciences, but also by the increasing number of fields that are becoming 
data intensive.  Hence the view from outside HEP is of a growing role for tape in an 
optimized scientific data management hierarchy.

Estimating tape costs is challenging.  The smallest component is usually the tape 
media, followed in ascending order by the acquisition of the drive-plus-robot system, 
the maintenance of the drives and robots, and the highly skilled labor needed to 
operate arcane tape-data-management systems such as HPSS.  It is reasonable to 
expect that as tape consolidates its position in scientific computing, less 
labor intensive approaches to its integration in the storage hierarchy will appear, 
not least due to development efforts at NCSA and other major scientific computing 
centers that see a clear need for this integration.

Tape technology availability has, for many decades, been limited by marketing considerations 
more than by technology itself.  Tape is very far from its fundamental physical 
limitations and suffers mainly from the existence of very few -- approximately 
two today -- major drive manufacturers who can see no reason to compete with 
themselves.  The technology could deliver twice the capacity per dollar every 
18 months, but it seems reasonable to expect a doubling time of around three 
years for the marketed products.

\subsubsection{Rotating Disks}
The 30-year run of exponential growth in capacity per dollar is almost certainly over.  
The factors responsible are both market related and technology related.

The consumer market for rotating disks is now declining.  An increasing number of 
consumer devices -- for example most HEP laptops -- are now being bought with flash 
memory taking over the historical role of disk.  One consequence is that an appealing 
and for-a-short-time successful HEP storage strategy -- buy the cheapest, slowest, 
largest consumer-market disks and hide them behind a layer of faster enterprise disks 
or solid-state-disks -- no longer works as well as it did.

In the enterprise-disk world, devices are loosely classified by their interfaces. 
SATA disks with low rotational speed and maximum capacity are used for less demanding 
“nearline” applications and much more expensive SAS disks are used for the more 
demanding “online” applications.  The cost differences are not really related to the 
SATA or SAS interfaces, but rather to properties such as rotational speed, quality of 
mechanical engineering, device monitoring probes,  and automatic recovery 
firmware.   The market for the most 
expensive disk capacity, currently provided by 2.5 inch 15k rpm drives used 
primarily for database applications, is already threatened by solid-state 
storage with much lower latency. 

\begin{figure}[h]
\centering
\includegraphics[width=0.9\textwidth]{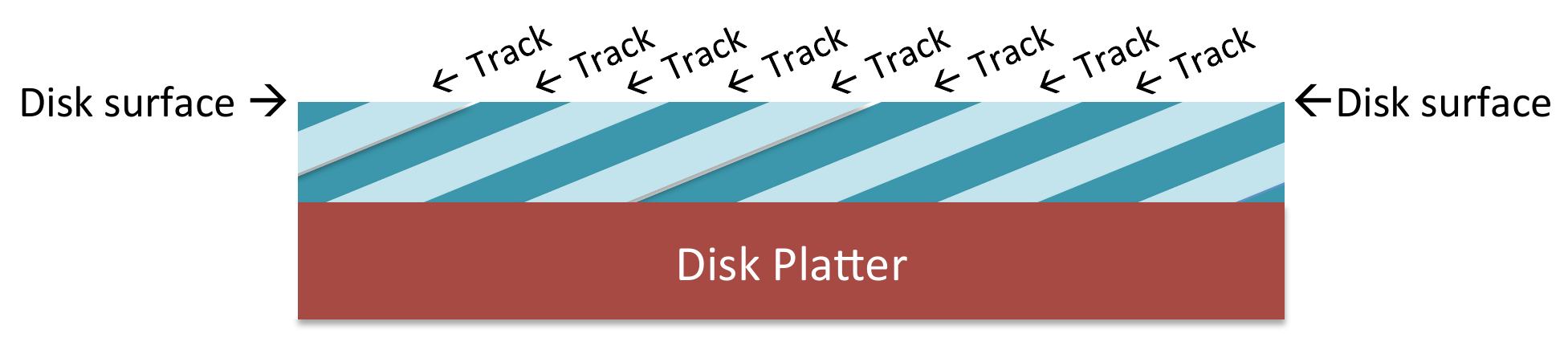}
\caption{Disk track layout produced by shingled recording.  Tracks are written from left to right in the diagram by a head that produces a magnetized volume with a triangular cross section. Individual tracks cannot be modified or erased}
\label{fig:shingled-recording}
\end{figure}

The technological problems faced by magnetic disk are easy to understand. The area 
of the bits written to current 4 terabyte drives is close to the limit of magnetic 
stability.  The density could be raised if more atoms could be involved by writing 
bits that used more of the magnetic material below the surface of the disk platter, 
and/or by using a higher coercivity material that would allow smaller bit sizes.  
Technological developments along these directions have been in the works for some 
years, but according to industry experts, none will be advanced enough to bring the 
next jump in disk capacity to market much before the end of the decade.  The prime 
``use-a-larger-volume-of-magnetic material'' approach is shingled recording, 
conceptually illustrated in Figure \ref{fig:shingled-recording}. In this approach a specially shaped write 
head lays down tracks with a triangular cross section, which become a set of 
``shingles'' as successive tracks are written.  Each shingle goes deep into the 
material, so can be stable even if it is much narrower than current tracks.  
It is impossible to re-write sectors or single tracks with this approach.  
To change any information on the disk, a large block of tracks must be erased 
and re-written.  HEP would have no problem with such a device, but its success 
in the wider marketplace could be problematic.  The ``high-coercivity'' approaches 
include Heat-Assisted Magnetic Recording (HAMR) using a uniform surface with higher 
coercivity and ``bit patterned recording'' using lithography to lay down a circular 
pattern of bits that are close together but physically isolated from each other.

The next ten years of disk evolution is thus highly uncertain.  It would be 
prudent and probably correct to assume that the doubling time is now around four 
years.  The impact of this will be considerable.  The logic that made it 
cost-effective to replace storage after as few as three years (approximately 
twice the doubling time) might now argue for keeping disks eight years or more, 
if the disks were sufficiently reliable.  Disks may also last longer if most of 
the access traffic can be fielded by a substantial solid-state storage cache as 
part of the overall storage hierarchy. There will be strong market pressures 
favoring, long lasting, low power, physically dense, and therefore very heavy 
disk systems.

\subsubsection{Solid-State Storage}
The certainty of an increasing role for solid-state storage is suggested by 
the twenty-year failure of rotating disk to provide more accesses per second 
per dollar.  In terms of this metric, today's solid-state storage provides 
vastly better value than rotating disk.  The solid state storage being used 
in (exploratory) production at HEP computing centers costs about ten times as 
much per unit capacity as rotating disk.  Reducing a rotating disk purchase 
by 10\% and spending the money on solid-state cache is cost neutral and can 
improve physics analysis throughput.

In this caching mode, solid-state storage is certain to play an increasing but not 
dominant role in HEP, and perhaps an even larger role in the commercial world.  
The woes of rotating disk described above are likely to allow solid-state -- 
specifically flash-memory-based -- storage to lower its cost relative to disk 
and gain market share.  It seems plausible that the cost differential would 
go down from around a factor ten now to around a factor 3 within ten years.  
Industry experts do not expect current solid-state technologies to “kill” disk 
in the foreseeable future, not least because it is almost inconceivable to 
create the large number of multi-billion dollar ``fabs'' -- chip fabrication 
facilities -- that would be needed to displace the world's many exabytes of disk.

\subsection{Storage Middleware and Data Management}
With the notable exception of the adoption by several HEP sites of the 
semi-commercial HPSS (High Performance Storage System)\cite{HPSS:url} to manage tape data, 
single-site storage middleware in HEP has been almost entirely home made.  In the 
late 1990s, it seemed that it would be an achievement to limit HEP's invention 
of tape data management systems to only one per lab.

HEP's middleware stacks have proved quite successful, for example the seamless 
integration for BaBar at SLAC of petabyte (HPSS) tape storage and tens of terabytes 
of disks distributed over 50 servers, but these successes have not translated into 
wide HEP adoption, let alone interest by other sciences or industry. 

In the wider scientific world, middleware such as Globus Online (GO)\cite{GlobusOnline:url}, which is 
a data transfer mechanism that has nice retry and a graphical display, is 
currently the norm.  These tools continue to grow and change, but are still 
based on the gridftp protocol.  There are other tools for data transfer, but 
it still has to be managed by the application.  Tools are also emerging that 
that transfer data at the file system level.  

Scalable file systems such as Lustre\cite{Lustre:url} and GPFS\cite{GPFS:url} connect 
the file system with the archive or nearline environment so there is seamless
access, from the users prospective, to all files.  The data-managed file 
system means that the inode remains in the file system with a stub of the first 
blocks of the data, or maybe nothing of the file at all, and the I/O is captured 
until the data returns to the local disk of the file system server.   
The data can be recalled from any ftp archive as long as permissions and 
security have been prearranged.   This builds basically an endless file system.   
Connecting these large endless file systems is then next on the horizon.  

In the next few years some test sites will be using GO-Storage 
to connect these large managed file systems together. The meta-data will be 
separate, in a replicated environment, and the data could be anywhere in the 
world.  As data is retrieved from locations, it moves closer and locally 
depending on the application requirements.   These methods all have small 
latencies added and depending on the actual location of the data, the latency could 
be great especially if it's on tape somewhere.   If these efforts yield systems and services 
that are widely adopted by science they must become serious candidates for 
meeting HEP needs in the future.

Even so, at the current time there are no commercial or widely used open-source 
offerings meeting the HEP needs for worldwide data and workflow management.  
The absence of commercial offerings is probably largely due to HEP's need to 
integrate tens to hundreds of autonomous computer centers, having many different 
funding sources, many different technologies, and a varying level of affiliation, 
from loose to none at all, with any particular HEP project.  Multinational commercial 
entities are generally orders of magnitude more coherent in their technologies and 
management.  The wider world of collaborative science does face many of the HEP 
challenges, but HEP is still at the bleeding/leading edge in terms of overall 
complexity.

The commercial world of distributed data-intensive applications should not be 
ignored -- individual technologies and {\it de-facto} standards relevant to HEP are 
almost certain to appear.  The wider scientific world must also not be ignored.  
HEP will be able to bring benefit, reap credit and attract non-HEP funding by 
helping adapt some of its most successful distributed computing technologies 
for other sciences and smaller HEP activities. Beyond this, HEP can benefit 
from participating in the development of widely useful distributed computing 
tools, and should studiously avoid the arrogance that could postpone the adoption 
of tools developed outside HEP to serve the exponential growth of data-intensive science.

A major stimulus to obey the exhortations above must be the labor intensive aspects, 
in both ongoing development and operations, of the distributed computing systems 
used in HEP today.

\section{Data, Software and Physics Preservation}
It is widely recognized that “knowledge preservation” in all frontiers is becoming 
increasingly important, or even vital, as the size of data samples and the time 
scale of experiments increase.  Here, ``knowledge preservation'' includes the preservation 
of data, the accompanying software, the capability to apply the software to the data 
in a reproducible way, and a record of how the software was applied to obtain a given 
result.  There are several motivations for effort in this area:
\begin{itemize}
\item
Enlightened self-interest: foresight and infrastructure is required to be able to 
re-use data and reproduce physics results several years later, even within a single 
experiment.  The LHC experiments are already seeing limitations in this area, even 
after only three years of data-taking.  Solving this problem in a generic way will 
enable most aspects of the issues behind knowledge preservation to be addressed in 
a manner that will allow many different experiments across the different frontiers 
to adopt a common solution.
\item
Outreach: limited access to some data and the accompanying documentation provides one 
means of preserving the ensemble of data and software.  Resources such as Rivet\cite{Rivet:url}, 
RECAST\cite{RECAST}, and the extensions to HepDB provide a means of encapsulating the structure 
of a given analysis in a manner that allows easy communication between experts and 
relatively simple re-use.  More broad outreach efforts, such as those supported by 
nearly all Energy Frontier experiments, have also developed simple yet powerful 
``public-friendly'' interfaces to simplified experimental data and analysis tools.
\item
Mandate: stewardship of the public investment in fundamental research is very important. 
(Have to write this carefully) While this is recognized by many of the experiments, 
HEP as a field has been reluctant to make large datasets public due their inherent 
complexity and the perceived difficulty in their interpretation by the non-expert.  
However, there is increasing pressure to make some the data available to the public 
in some form.  Doing this in a manner that is sustainable and does not require immense 
resources poses many difficult problems which would have to be solved if this pressure 
turns into a mandate from the funding agencies.
\end{itemize}

\subsection{Current Efforts in Knowledge Preservation}
Worldwide, and across many disciplines, the past couple of years have seen widespread 
recognition that Data/Knowledge preservation is critical to the long-term exploitation 
of any science that qualifies as ``Big Data''.  This broad interest has spawned a 
dizzying array of national and international disciplinary and cross-disciplinary 
efforts to attempt to address the related issues.  Up to now, many of these have 
focused solely on providing a means of storage and indexing of datasets.  This is 
insufficient for many of the particle and astrophysics experiments, who require a 
means of recording processing workflows as well.  Indeed, particle experiments 
probably have the most stringent and complex requirements for knowledge 
preservation, driven primarily by the complexities of the datasets and the amount 
of processing required to achieve a physics result.

The DPHEP effort\cite{DPHEP}, a Study Group for Data Preservation in High Energy Physics under the auspices of the International Committee on Future Accelerator (ICFA), has produced a study outlining the current state of data preservation within HEP, including an extensive overview of other disciplines which will not be reproduced here. They suggest a series of guidelines for HEP data preservation efforts, as well as a framework for global coordination.  Their conclusions include a recognition of the scientific potential for data re-use, especially the desirability to preserve full analysis capability.  They also emphasize the urgency required to begin and sustain global, coordinated data preservation efforts.  

Among colliders whose data-taking period has ended, two currently have advanced efforts in data preservation and access.  The BaBar Long Term Data Access archival system\cite{BaBar} is comprised of several racks of compute nodes and data and database servers, all completely walled-off from the host laboratory.  The deployment of virtual machines ensures a stable, non-evolving software environment.  This system is designed to provide active analysis use through 2018, at which point only the legacy data would be preserved.  At DESY, a combined effort is underway to maintain sustained access to the HERA datasets of ZEUS, H1, and HERMES.  A cross-HERA virtual environment based on current grid infrastructure that will allow new production of simulated events already exists; each of the experiments has slightly different plans for the long-term archiving and access to their data and the corresponding documentation.  A general software/archival validation tool is also under development at DESY\cite{DESY}; this could become a general component in any long-term software and data preservation effort.
Planning at the Tevatron is underway\cite{Tevatron}. The Tevatron experiments finished data collection at the end of September, 2011. Unlike at the LHC, very little thought was given to data preservation and re-use during Tevatron running, which will make the preservation of the data especially challenging. 

At the LHC, K. Cranmer’s RECAST project is notable for its advanced state of deployment and its differences from traditional archival methods.  RECAST is a computational framework designed to encapsulate the “institutional knowledge” about a particular analysis in a way that allows outsiders to query the dataset for a certain physics result.  The analysis methods, the simulated events, and the results from data are stored in a manner that can be used to probe the presence or absence of new physical processes that were not considered at the time of the analysis.  So, the analysis algorithms and the dataset are fixed, but the results and constraints on new physics models can be explored.  

The LHC experiments are currently considering data preservation and public access policies.  CMS and LHCb have approved policies on public access to data and on the prospects of releasing processed data to the public.

The Data and Software Preservation for Open Science (DASPOS) project, funded by the National Science Foundation, has established a collaboration tying together US physicists from the CMS and ATLAS experiments at the LHC and the Tevatron experiments, scientists from other disciplines, and experts in digital preservation, heterogeneous high-throughput storage systems, large-scale computing systems, and grid access and infrastructure. The DASPOS activities are connected with the DPHEP coordination effort, the experimental collaborations, and other related multi-disciplinary projects in Europe, Asia, and the US. Together, this group represents the US in international efforts, and acts as a coordinating point of contact, a partner in dialogue, and a technological consort.  The DASPOS project represents an initial exploration of the key technical problems that must be solved to provide appropriate data, software, and algorithmic preservation for HEP and will address the importance of preserving the contexts necessary to understand, trust, and re-use data. It will attempt to create a template architecture for
knowledge preservation in data-based science, focused on HEP, but representing broader consultation and applicability.  The 
activities of the project comprise a series of workshops aimed at fact-finding and cross-disciplinary exploration coupled with a technical demonstration of a knowledge preservation infrastructure.

While DPHEP serves as a central forum for discussion of Data/Knowledge preservation in high energy physics, there are many other efforts in other scientific fields that are attempting to grapple with similar problems.  This is one area where central guidance and coordination of expertise across funding agencies and scientific entities would be very welcome.

\section{HEP Outlook and Recommendations}
\label{sec:cpfi5-hep-outlook}
\subsection{Impact of the Technology Outlook on HEP}
\begin{figure}[h]
\centering
\includegraphics[width=0.7\textwidth]{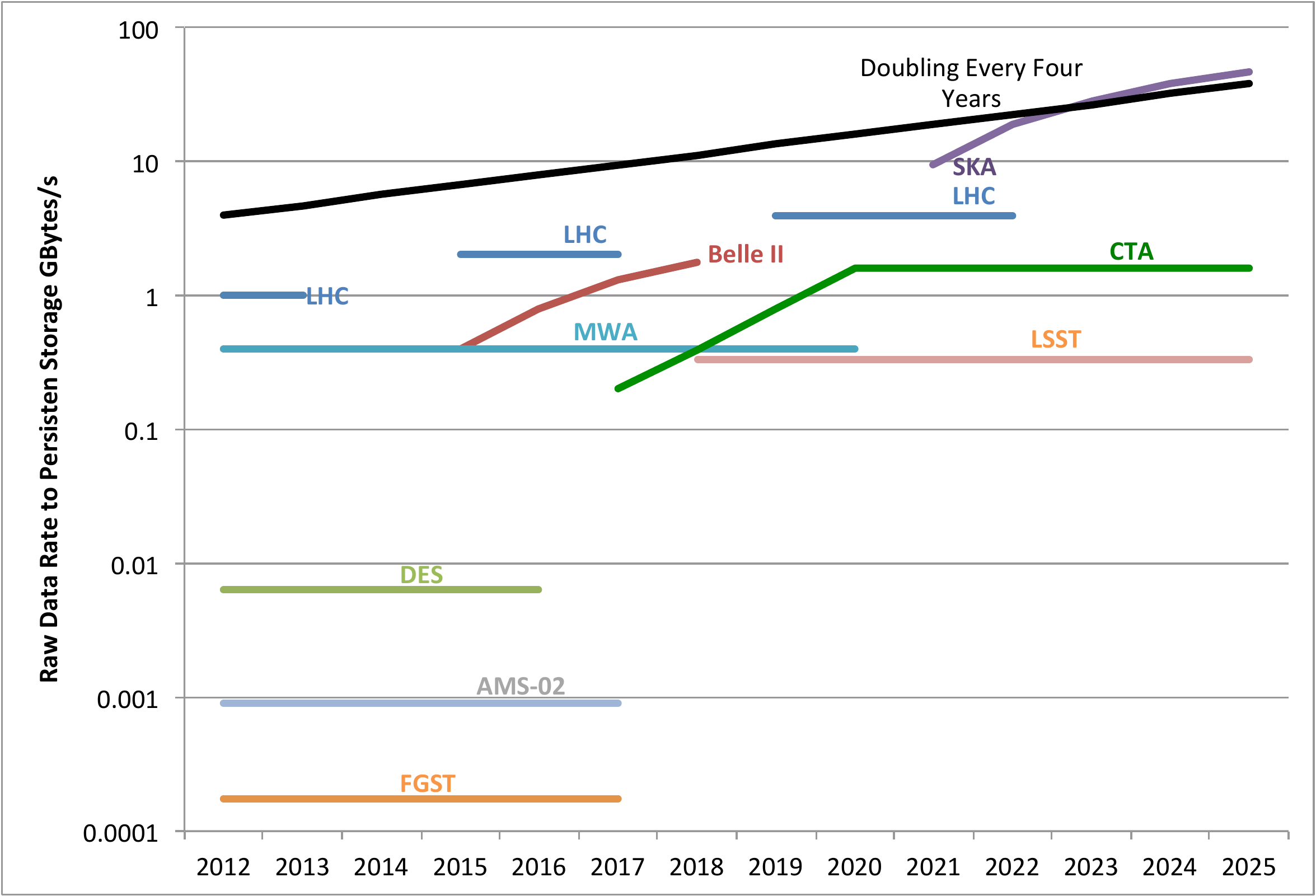}
\caption{Raw data rates of HEP experimental/observational programs in comparison with likely capacity/unit price evolution}
\label{fig:storage-evolution}
\end{figure}

The coming decade and beyond will see experiments and observations at several 
frontiers straining against the limits of data management and storage technology.  
Storage technology is likely to evolve in capacity/cost more slowly than in the 
last decades, making it ever more important that the role of storage is carefully 
optimized with respect to other costly elements of the scientific programs.  
Figure \ref{fig:storage-evolution} attempts to illustrate this challenge by comparing the data rates 
expected from the most challenging activities with the disk technology evolution 
predicted earlier in this report.  

As outlined earlier, the rate of writing raw events to persistent storage for the 
LHC experiments is substantially influenced by the cost of storage and can be 
expected to evolve at a similar rate.
It is likely that other activities with petabyte/s rates for raw data from 
front-end devices will be compelled to employ real-time data processing and 
selective rejection in order to reach persistent rates that are affordable 
in the context of each activity.

At a more detailed level, the roles of tape, rotating disk and solid-state 
storage will evolve:
\begin{itemize}
\item
Tape will fall slightly in relative cost should and play a more important role 
than, for example, in the LHC experiments today where it is arguably underused.
\item
Rotating Disk, will not improve in capacity/cost as rapidly as tape or solid-state 
storage and for the next decade may take around four years to double capacity/cost.  
As a result, it could be cost-effective to wait about eight years before replacing 
disks that are not giving trouble, provided the space, power and cooling are available to make this possible.
\item
Solid-state storage will not kill rotating disk in the enterprise and data-intensive 
science markets, but it will become relatively cheaper, perhaps by a factor of three, 
and will merit full consideration in the overall optimization of HEP computing 
environments, where it offers the promise of efficient sparse or random access 
to HEP data.  To make effective use of the still small affordable quantities of 
solid state storage, application-aware approaches to caching data on such storage 
appear essential.
\end{itemize}

The relative rates of cost evolution of tape, rotating disk, solid-state storage, 
networks and CPU are clearly uncertain.  However, it is certain that HEP computing
models will need to adapt to make the most efficient use of all these elements.  
The best way to adapt would be to make the implementation of the models 
intrinsically adjustable to serve a wide range of situations. The now old, 
but never really implemented, concept of ``virtual data'' would go a long way to 
making HEP computing dynamically adaptable to wide range of relative storage and 
CPU costs.  In a virtual data system, all derived (or simulated) data products 
begin existence solely as the rigorously complete recipes for creating them.  
These virtual data products are then instantiated or replicated based on real 
or algorithmically anticipated demand.  The physical instances are retained 
based on anticipated future use and the relative cost of storage versus re-creation.  
HEP has often appeared to be close to developing the rigorous provenance databases 
needed to implement virtual data and a full virtual data implementation may be in 
reach of the LHC experiments for Run 3. 

\section{Findings and Recommendations}
\label{sec:cpfi5-findings}
{\bf CpF15 Finding 1a:} The largest HEP experiments have developed, and are 
improving functional distributed data and workflow management systems meeting 
their needs. These systems are expensive to develop and operate and are thus 
rarely appropriate for smaller experiments.

{\bf CpFI5 Finding 1b:} HEP currently benefits from, but can also be 
constrained by, the highly successful ROOT features supporting reading 
and writing of persistent data. No other major scientific field uses 
ROOT or appears interested in it. Major developments in persistency 
technology will be required to take advantage of storage hardware on the 
timescale of LHC Run 3.

{\bf CpFI5 Recommendation 1:} HEP should maintain and promote a vision of the future 
in which fully functional and low-operational-cost distributed computing and 
persistency management is supported by software that is widely used in 
data-intensive science.  To this end, developments in industry and the wider 
science community should be monitored actively, HEP should work with the wider 
science and computer science community to export and adapt HEP technologies and 
vice-versa. In distributed computing, HEP should organize itself to significantly 
reduce the number of diverse approaches and provide the benefits of ideas and 
software developed in the largest experiments to other activities where they are needed.

{\bf CpFI5 Finding 2a:} Rotating disk storage will suffer a marked slowdown in the 
evolution of capacity/cost.  This may be the largest perturbation of HEP computing 
models that must attempt to optimize the roles of tape, rotating disk, solid-state 
storage, networking and CPU.

{\bf CpFI5:} Finding 2b: Many of the components required to support virtual data 
already exist in the data and workflow management software of the largest experiments.  
The rigorous provenance recording required to support the virtual data concept would 
also benefit data preservation.

{\bf CpFI5 Recommendation 2:} Computing model implementations should be flexible 
enough to adapt to a wide range of relative costs of the key elements of HEP 
computing.  In preparing for Run 3, the LHC program should seriously consider 
virtual data as a way to accommodate scenarios where storage for derived and 
simulated data becomes relatively very costly.

{\bf CpFI5 Recommendation 3:} All experiments across all frontiers need 
infrastructure that will allow scientists to store, catalogue, access, and 
reprocess datasets years after the original physics results are produced. 
The inherent similarity of the requirements across experiments and disciplines 
call for a coordinated investment in common infrastructure to enable easy 
access and adoption of best practices in knowledge preservation.  Solutions 
should be developed that meet the needs of the particle and astrophysics 
communities before widespread release of data to the public can be expected 
or mandated.


\end{document}